\documentclass[12pt]{article}
\usepackage{epsfig}
\usepackage{amssymb,graphicx}
\def\be{\begin{equation}}
\def\ee{\end{equation}}
\def\ba{\begin{array}{c}}
\def\ea{\end{array}}

\def\ben{\[}
\def\een{\]}

\newcommand{\bea}{\begin{eqnarray}}
\newcommand{\eea}{\end{eqnarray}}

\begin{document}


\begin{center}

{\Large \bf {


Morse potential, symmetric Morse potential and bracketed bound-state
energies

 }}

\vspace{13mm}


 {\bf Miloslav Znojil}

 \vspace{3mm}
Nuclear Physics Institute ASCR, Hlavn\'{\i} 130, 250 68 \v{R}e\v{z},
Czech Republic

 e-mail:
  znojil@ujf.cas.cz

\vspace{3mm}


\end{center}



\section*{Abstract}

For the needs of non-perturbative quantum theory an upgraded concept
of solvability is proposed. In a broader methodical context the
innovation involves Schr\"{o}dinger equations which are piece-wise
analytic and piece-wise solvable in terms of special (in our
illustrative example, Whittaker) functions. In a practical
implementation of our symbolic-manipulation-based approach we work
with a non-analyticity in the origin. A persuasive advantage is then
found in the both-sidedness of our iterative localization of the
energies.

\subsection*{Keywords:}

quantum bound states; special functions; Morse potential;
symmetrized Morse potential; upper and lower energy estimates;
computer-assisted symbolic manipulations;

\subsection*{PACS}
.

02.30.Gp – Special functions

03.65.Ge – Solutions of wave equations: bound states

11.30.Pb -  Supersymmetry

%
%

\newpage

\section{Morse potential and the molecular-spectrum paradox
\label{Introduction}}

\begin{figure}[h]                     
\begin{center}                         
\epsfig{file=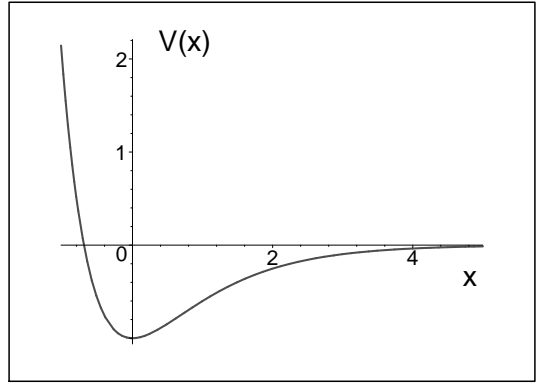,angle=270,width=0.4\textwidth}
\end{center}                         
\vspace{-2mm}\caption{Function (\ref{Morsepo}) at
$\alpha=\gamma_1=\gamma_2=1$.
 \label{picee3wwww}}
\end{figure}

Morse potential \cite{Morse}
 \be
 V(x)=V_{(Morse)}(x) =-2\,\gamma^2_1\,e^{-\alpha\,x}+\gamma^2_2\,
 e^{-2\alpha\,x}
 \label{Morsepo}
 \ee
say, in its special two-parametric form with
$\gamma_1=\gamma_2=\gamma$ (cf. Fig.~\ref{picee3wwww}) belongs to
the family of shape-invariant interactions which make the
one-dimensional Schr\"{o}dinger bound-state problem
 \be
 -\, \frac{{\rm d}^2}{{\rm d} x^2} \psi_n(x)
 + V_{}(x) \psi(x)= E_n\,
 \psi_n(x)\,
  ,\ \ \ \ \ \psi_n(x)  \in L^2(-\infty,\infty)\,
   \label{SEx}
  \ee
solvable, in closed form, in terms of classical orthogonal
polynomials. Such an exactly solvable (ES) family (which is
intimately related to the supersymmetric model building, cf.
\cite{Cooper}) is not too large. This means that potential
(\ref{Morsepo}) (for which the polynomial part of the exact wave
functions $\psi_n(x)$ are Laguerre polynomials) is, in the context
of mathematics of differential equations, exceptional.

Remarkably, the same potential is also exceptional from the point of
view of applied quantum theory and, in particular, of molecular
physics. Indeed, after an appropriate selection of parameters, model
(\ref{Morsepo}) offers one of the best numerical fits to the
measured vibrational spectra of diatomic molecules \cite{Walger}.
Alas, strictly speaking, the {\em exact\,} solvability of the
one-dimensional Schr\"{o}dinger Eq.~(\ref{SEx}) with
$V_{}(x)=V_{(Morse)}(x)$ only serves, for the purposes of the fit,
as an {\em ``excellent approximation''} (see, e.g., pages 182-185 in
Ref.~\cite{Fluegge} for more details).

In the language of mathematics such an ``approximation paradox''
originates from the three-dimensional nature of the realistic
molecular Schr\"{o}dinger equation in which the Morse potential
itself only emerges in the partial-wave sequence of the so called
{radial} Schr\"{o}dinger equations
 \be
 -\, \frac{{\rm d}^2}{{\rm d} r^2} \psi_{n,\ell}(r)
 +\frac{\ell(\ell+1)}{r^2} \psi_{n,\ell}(r)
 + V_{(\ell)}(r)  \psi_{n,\ell}(r)= E_{n,\ell}\,
  \psi_{n,\ell}(r)\,
  ,\ \ \ \ \ n,\ell = 0, 1, \ldots\,
   \label{esSEx}
  \ee
where \cite{Walger}
 \be
 V_{(\ell)}(r)= V_{(Morse)}(r-d)-\frac{\ell(\ell+1)}{r^2}\,.
 \label{merela}
 \ee
Thus, the realistic molecular wave functions $\psi_{n,\ell}(r)$ only
live on the {\em half-lines} of the radial coordinate $ r \in
(0,\infty)$ and vanish in the limit $r \to \infty$. Naturally, one
has to impose the conventional boundary condition in the origin
\cite{Comment},
 \be
  \psi_{n,\ell}(0)=0\,.
  \label{bc0}
  \ee
Even if we accept the trivial regularization (\ref{merela}) of the
strongly singular centrifugal term in Eq.~(\ref{esSEx}), we reveal
that the power-series ansatz for wave functions {\it does not
terminate} and  that it {\it cannot} degenerate to a Laguerre
polynomial. In other words, the molecular wave functions
$\psi_{n,\ell}(r)$ of Eq.~(\ref{esSEx}) cannot coincide with the
exact solutions $\psi_{n}(x)$ of Eq.~(\ref{SEx}). Similarly, the
``realistic'' molecular bound-state spectrum of energies
$E_{n,\ell}$ may only be determined via certain real, {\em purely
numerical\,} roots of transcendental Eq.~({\ref{bc0}). Thus, the
spectrum of the Morse-potential model of a molecule will {\it not}
be given by the exact analytic formula, say, of Table Nr. 4.1 in
Ref.~\cite{Cooper}.

In what follows we intend to discuss the latter paradox and we shall
draw, from this discussion, a few nontrivial methodical as well as
model-building consequences.

\section{Resolution of the paradox and the new, symmetrized Morse potentials}

{\it A priori}, the latter conclusions might appear rather
discouraging. Fortunately, the errors caused by {\it both} of the
currently accepted approximations (\ref{bc0}) and (\ref{merela})
prove entirely negligible even in the most unfavorable cases in
practice (cf., e.g., a compact outline of the problem in
\cite{Fluegge}, p. 184). Thus, the enormous phenomenological appeal
of the exactly solvable Morse potential in molecular physics may be
perceived as born out of a serendipitious practical
indistinguishability between the purely numerical values of the
``exact'' roots of the ``realistic'' secular Eq.~(\ref{bc0}) and the
non-numerical, purely analytic values as derived from the
``idealized'', ES Schr\"{o}dinger Eq.~(\ref{SEx}).

Recently, a new, innovative version of the closeness between the
numerical transcendental-secular-equation roots and their efficient
analytic approximations has been found and used by Ishkhanyan
\cite{Ishkh}. He conjectured that such a closeness (i.e., an {\em
``excellence of approximation''}) might emerge in multiple
non-standard quantum eigenvalue problems on half-line. He
demonstrated the feasibility of such a project via its application
to the inverse square root potential $V(r) = V_0/\sqrt{r}$. He
illustrated the practical appeal and the user-friendliness of such a
new model-building strategy by showing that the construction may
prove successful in Morse-resembling cases in which the general
differential-equation solutions $\psi_{n,\ell}(r)$ entering
differential Eq.~(\ref{esSEx}) remain proportional to a suitable,
asymptotically correct confluent hypergeometric special function.

Recently \cite{qes} we imagined that besides its usual connection
with the half-line radial Schr\"{o}dinger equation, boundary
condition (\ref{bc0}) may be equally well interpreted as a matching
condition for all of the odd-parity bound states living on the whole
real line. In other words, one can move from radial half-line
Eq.~(\ref{esSEx}) to the alternative, one-dimensional bound-state
problem (\ref{SEx}) in which the potential is artificially
``symmetrized'', i.e., such that $V(x) = V(-x)$. In this manner, in
particular, one gets, ``free of charge'', all of the odd-parity
bound states in the left-right-symmetrized Morse potential
 \be
 V_{(sym.)}(x,d)=\left \{
 \begin{array}{ll}
  V_{(Morse)}(-d+x),& x>0,\\
  V_{(Morse)}(-d-x),&x<0\,,
 \ea
 \right . \ \ \ \
 d>0\,
  \label{dolenhol}
 \ee
(cf. Fig.~\ref{ee3wwww}). In other words, after one deliberately
replaces the ``strongly asymmetric'' function $V_{(Morse)}(x)$ of
Fig.~\ref{picee3wwww} by its full-line-interaction alternative of
Fig.~\ref{ee3wwww}, the odd-parity spectrum of energies (given by
the roots of Eq.~(\ref{bc0})) remains unchanged.

\begin{figure}[h]                     
\begin{center}                         
\epsfig{file=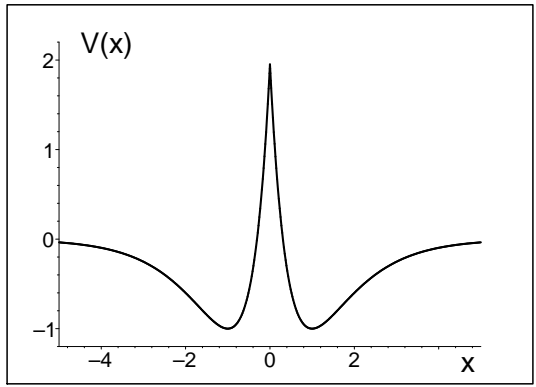,angle=270,width=0.4\textwidth}
\end{center}                         
\vspace{-2mm}\caption{Symmetrized descendant (\ref{dolenhol}) of
potential of Fig.~\ref{picee3wwww} with shift $d=1$.
 \label{ee3wwww}}
\end{figure}


The ``missing'', even-parity rest of the spectrum of the new
solvable model (\ref{SEx}) + (\ref{dolenhol}) will be obtained after
the replacement of Eq.~(\ref{bc0}) by its alternative
 \be
  \lim_{x\to 0}\frac{d}{dx}\,\psi_{n,\ell}(x)=0\,,
  \label{bc1}
  \ee
i.e., {\it mutatis mutandis}, by the virtually equally well defined
transcendental (i.e., special-function-based,
confluent-hypergeometric-series-based) secular equation again.

\section{Bound states}

At the larger shifts $d$ the barrier as sampled by
Fig.~\ref{ee3wwww} becomes very high so that the even- and
odd-parity bound states will remain almost degenerate. More
interesting spectra will only  be obtained at the not too large,
positive (or even negative, barrier-free!) values of the shift $d$.

\subsection{Analytic considerations}

On the basis of the practical experience as made by molecular
physicists the polynomially solvable Morse-interaction problem
(\ref{Morsepo}) + (\ref{SEx}) offers a very good approximation to
the non-polynomial wave functions of radial Eq.~(\ref{esSEx}). While
the non-polynomiality caused by the matching condition
$\psi_{n,0}(0) = 0$ in the origin remained negligible, the advantage
of having the approximate solutions in a safely analytic,
non-numerical, confluent-hypergeometric-function form was decisive.

After transition to our present double-well model (\ref{SEx}) +
(\ref{dolenhol}}) we still can use  the same formal advantages
including both the quality of the existing non-numerical
approximations of functions $\psi(x)$ (or bound-state energies) at
$d \gg 0$, {\em and} the tractability of the {\em general}
confluent-hypergeometric solutions at a fixed shift $d$ and at a
variable energy $E_{trial}=-k^2_{trial}$. In the intermediate region
of $d$s the high precision of the usual approximations (based on the
Stirling formula, cf., e.g., equation Nr. (70.15) in monograph
\cite{Fluegge}) will worsen so that it should certainly be
systematically amended via higher-order corrections.

It is worth adding that the present introduction of an anomalous
non-analyticity of the potential in the origin enables us to replace
the analytic but asymmetric potential (\ref{Morsepo}) by its
symmetrized but non-analytic version (\ref{dolenhol}) so that at
least some of the formal advantages of having the traditional
analytic Morse potential (as sampled, e.g., in Ref.~\cite{MorsePT})
may be lost.  In parallel, the new interaction model
(\ref{dolenhol}) also possesses specific merits. For example, it
provides a double-well shape (cf. Fig.~\ref{ee3wwww}) giving rise to
the tunneling. The internal barrier is tuneable, of a freely
variable thickness.

Naturally, the similar appealing descriptive features may be found
in multiple other models, the list of which would range from the
purely numerical Mexican-hat-shaped quartic-polynomial potentials
\cite{Goldstone} and from the various semi-numerical square wells
with two minima \cite{2sqw} up to the exactly solvable extreme of a
pair of attractive delta-function potentials \cite{svitou}.
Nevertheless, a combination of the strong central repulsion with a
finite height of the barrier seems to be a specific an unique
phenomenological aspect of Eq.~(\ref{dolenhol}).

\subsection{Numerical considerations}

The most common, computation-oriented distinctive feature of model
(\ref{dolenhol}) with $d \gg 0$ lies, as we already emphasized, in
the amazing numerical efficiency of the approximate replacement of
the numerically difficult $r_0=0$ boundary condition
$\psi_{n,\ell}(r_0)=0$ in realistic Eq.~(\ref{esSEx}) by the full
line problem (\ref{SEx}) with the Dirichlet boundary condition
shifted to $r_0 \to -\infty$. After one adds the alternative
boundary condition (\ref{bc1}), one must compute higher order
corrections so that the next-to-degenerate problem becomes
mathematically challenging again.

\begin{figure}[h]                     
\begin{center}                         
\epsfig{file=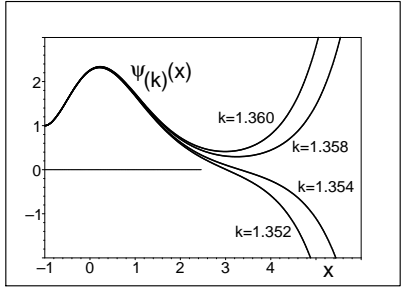,angle=270,width=0.3\textwidth}
\end{center}                         
\vspace{-2mm}\caption{The influence of errors in energy $E=-k^2$
upon the asymptotics of the (right half of the) trial-and-error
ground-state regular wave functions. Potential of
Eq.~(\ref{dolenhol}) is used with parameters $d=1$, $\alpha=1$,
$\gamma=1.8$. Normalization $\psi(0)=1$ and $\psi'(0)=0$. }
 \label{srov2}
\end{figure}

Interaction  (\ref{dolenhol}) may be characterized by the most
common specific confluent-hypergeometric versions of wave functions
which are asymptotically correct (i.e., vanishing) and enter the
physical boundary conditions at $r=0$. In similar models, such an
approach is most common \cite{Ishkh}. Nevertheless, there also
exists an alternative approach in which the determination of the
bound-state energies would be based on the much less common (often
called ``regular'') specific confluent-hypergeometric versions of
the wave functions $\psi^{(reg.)}(x,E)$ for which we {\em
guarantee}, using analytic means, the correct behavior in the origin
(cf. Eqs.~(\ref{bc0}) and (\ref{bc1})) {\em in advance}, i.e., at
{\em any} energy $E$.

\begin{figure}[h]                     
\begin{center}                         
\epsfig{file=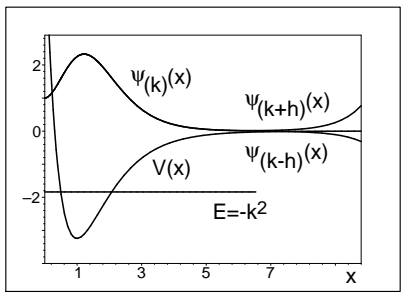,angle=270,width=0.3\textwidth}
\end{center}                         
\vspace{-2mm}\caption{ Right half of the ground-state
$\psi(x)=\psi(-x)$ with $\psi(0)=1$ in potential $V(x)=V(-x)$ of
Eq.~(\ref{dolenhol}) at energy $E=-k^2$, $k=1.355765$ and parameters
$d=1$, $\alpha=1$ and $\gamma=1.8$. The influence of error $h=
0.000005$ is only detected at very large $x \gtrsim 7$. }
 \label{bi2}
\end{figure}

One of the reasons why the use of the regular-function ansatzs
$\psi^{(reg.)}(x,E)$ is much less popular in practice is that their
explicit representation is usually rather complicated. Thus, in
their sample displayed in Fig.~\ref{srov2} the Whittaker-function
representation of the explicit formula for $\psi^{(reg.)}(x,E)$
which would be compatible with boundary condition (\ref{bc1}) had to
be generated via a symbolic-manipulation software. It remained too
lengthy to be displayed in print. This being said, the manipulations
which led to the production of Figs.~\ref{srov2} or \ref{bi2}
remained virtually trivial, fully comparable, say, with the
traditional algebraic manipulations with the elementary
trigonometric functions needed in the case of the piecewise-constant
square wells $V(x)$.

\begin{figure}[h]                     
\begin{center}                         
\epsfig{file=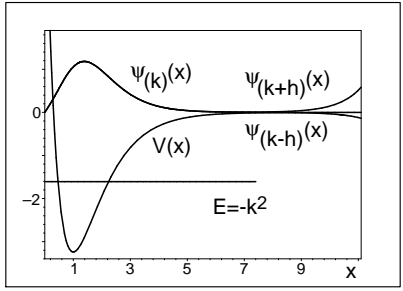,angle=270,width=0.3\textwidth}
\end{center}                         
\vspace{-2mm}\caption{ Right half of the lowest excited-state
$\psi(x)=-\psi(-x)$ in potential of Fig.~\ref{bi2} at energy
$E=-k^2$ with $k=1.268113$ and error bar $h=0.000003$ using
normalization $\psi'(0)=1$. }
 \label{hee2}
\end{figure}

Another, much more important specific merit of using regular
special-function solutions is also well illustrated by
Figs.~\ref{srov2}, \ref{bi2} and \ref{hee2}. Indeed, once we recall
the standard oscillation theorems \cite{Hille}, we may ``bracket''
the correct root $E_n=-k^2_n$ of the necessary and sufficient
asymptotic, regular-function-based  boundary-condition constraint
 \be
 L(E)=\lim_{x \to \infty} \psi^{(reg.)}(x,E)=0
 \label{8}
 \ee
by counting the nodal zeros, i.e., in a way which immediately
characterizes any preselected approximate value of the $n-$th-level
parameter $k_{trial}=\sqrt{-E_{trial}}$ as giving an upper or lower
bound (thus, in particular, Fig.~\ref{srov2} immediately informs us
that, with certainty, $1.354<k_{exact}<1.358$).

One of the reasons why the resulting symbolic-manipulation-assisted
version of the well known shooting method works much better than its
standard numerical predecessors is that the special-function form of
$\psi^{(reg.)}(x,E)$ provides the information with a fully
controlled and arbitrarily preselected precision. This feature is
illustrated by Figs.~\ref{bi2} and \ref{hee2} out of which one
reads, reliably, that for the potential in question we have
 \ben
 -(1.35577)^2
 < E_0<
 -(1.35576)^2
 \een
and
 \ben
 -(1.268116)^2
 < E_1<
 -(1.268110)^2
 \een
respectively.

\subsection{Methodical considerations}

Among the most popular simulations of quantum dynamics in one
dimension one finds various piecewise constant potentials. The main
advantage of such a choice of effective dynamics may be seen in the
related piecewise trigonometric form of components $
\psi^{(1,2)}_{(j)}(x)$ of the general, energy-dependent solutions of
Eq.~(\ref{SEx})
 \be
  \psi_{(j)}(x)=  C^{(1)}_{(j)} \psi^{(1)}_{(j)}(x)+
   C^{(2)}_{(j)} \psi^{(2)}_{(j)}(x)\,,
  \ \ \ \ \
   x  \in (a_j,a_{j+1})\,,
   \ \ \ \ \ j = 0, 1, \ldots, K
   \label{compo}
  \ee
such that
 \be
 -\infty=a_0<a_1<\ldots<a_K< a_{K+1}=\infty\,.
 \ee
In the experimental context, typically, one has to satisfy the
asymptotic boundary conditions
 \be
 \lim_{x\to \pm \infty}\psi_n(x)=0\,,\ \ \ \
 E=E_n<\min \, [V(a_0),V(a_{K+1})]
 \ee
in order to determine the observable bound-state energies $E=E_n$
and/or the wave functions $\psi(x)=\psi_n(x) \in L^2(\mathbb{R})$.
In such a setting, the pairwise matching of the logarithmic
derivatives of components (\ref{compo}) at the boundary-points $a_j$
couples the constants $C^{(i)}_{(j)}$. Then, the construction of
bound states degenerates to the search of roots of a certain
algebraic secular equation resembling Eq.~(\ref{8}). This indicates
the possible path of further generalizations of our present model.

\section{Summary}

The non-exact, approximative nature of the description of molecules
using Morse potentials is rarely emphasized in the literature. With
exceptions: The terminological inconsistency attracted our attention
in our older paper \cite{MorsePT}. We studied there the connections
between the half-axis of coordinates in (\ref{esSEx}) and the full
line of coordinates in (\ref{SEx}). Successfully we regularized
there the centrifugal singularity in a way based on an analytic
continuation of the wave functions to the complex plane of the
coordinates.

Now we  addressed the same problem from an opposite point of view.
Our main attention was paid to the one-dimensional phenomenological
scenario and to the obvious fact that the shape of $V_{(Morse)}(x-d)
\neq V_{(Morse)}(d-x)$ is so strongly asymmetric that its half-axis
reinterpretation (\ref{merela}) provided probably its only, albeit
approximative, truly useful contact with the three-dimensional
real-world experiments.

In molecular physics people weakened the concept of the exact
solvability when moving from the one-dimensional problem~(\ref{SEx})
to its more realistic three-dimensional generalization. In our paper
we accepted such a weakened concept and we described some of its
innovative model-building consequences. More explicitly, we returned
to the one-dimensional setting of Eq.~(\ref{SEx}) and considered a
new Morse-like potential (\ref{dolenhol}). The related
Schr\"{o}dinger bound-state problem was discussed in some detail.

In a way inspired by the common practice in molecular physics the
eigenvalue problem was characterized by the non-terminating
Taylor-series representations of wave functions. In both the
contexts of Eqs.~(\ref{SEx}) and (\ref{esSEx}) we considered the
general solutions $\psi_{n}(x)$ and $\psi_{n,0}(r)$ in their
respective explicit, confluent-hypergeometric-function forms.
Being encouraged by Ishkhanyan \cite{Ishkh} we decided to accept
the non-ES, non-polynomial, infinite-series special-function form of
wave functions as the form of bound states which can still be
welcome as non-numerical and exact. Our tests of the strategy
revealed an energy-bracketing behavior of the trial-and-error
eigenvalues so that we believe that it offers a robust algorithm for
practical computations.

In the context of physics we put main emphasis on the appeal of the
double-well shapes of potentials as sampled by Fig.~\ref{ee3wwww}.
Nevertheless, a more or less unmodified constructions of bound
states seem to remain applicable even if we  turn Fig.~\ref{ee3wwww}
upside down, yielding the single-well formula
 \be
 V(x)=V_{(single\ well)}(x)=\left \{
 \begin{array}{ll}
  -V_{(Morse)}(-d+x),& x>0\\
  -V_{(Morse)}(-d-x),&x<0
 \ea
 \right .\,,
 \ \ \ \ \ \ d > 0
  \label{horenhol}
 \ee
in which the central attractive part becomes accompanied by certain
external barriers. Although such a shape could, in principle,
generate low-lying resonant states, its study already lies beyond
the scope of our present paper.


\section*{Acknowledgements}

Inspiring correspondence with Artur Ishkhanyan is gratefully acknowledged. The work on the
project was supported by the Institutional Research Plan RVO61389005 and by the standard
GA\v{C}R Grant Nr. 16-22945S.


\end{document}